\documentclass[acmsmall,screen]{acmart}

\usepackage[utf8]{inputenc}
\usepackage{microtype}
\usepackage{textgreek}
\usepackage{fancyvrb}
\usepackage{multirow}
\usepackage{tabularx} 
\usepackage{colortbl}
\usepackage{longtable}
\usepackage{booktabs}
\usepackage{ifthen}
\usepackage{color}
\usepackage{bbding}
\usepackage{makecell}
\usepackage{threeparttable}
\usepackage{amsmath}
\usepackage{wasysym}
\usepackage{stfloats} 
\usepackage{xspace}
\usepackage[T1]{fontenc}
\usepackage{paralist}
\usepackage{enumerate}
\usepackage[linesnumbered,ruled,vlined]{algorithm2e}
\usepackage{array, multirow,graphicx}
\graphicspath{{Figures/}}
\usepackage{float}
\usepackage{balance}
\usepackage{tikz}
\usepackage{calc}
\usepackage{subfigure}
\usepackage{listings,amsfonts}
\usepackage{makecell}
\usepackage{xcolor}
\usepackage{listings}

\newcommand{\tool}{\textsc{Cve2PoC}\xspace}

\usepackage{tcolorbox}
\newtcolorbox{findingbox}{
  colback=gray!10,
  colframe=gray!10,
  boxrule=0pt,
  left=5pt,
  right=5pt,
  top=5pt,
  bottom=5pt
}
\newcommand{\finding}[2]{%
\begin{findingbox}
\textbf{Answer to RQ#1:} #2
\end{findingbox}
}

\definecolor{codegreen}{rgb}{0.2,0.5,0.2}
\definecolor{codeblue}{rgb}{0.1,0.3,0.6}
\definecolor{codegray}{rgb}{0.5,0.5,0.5}
\definecolor{codepurple}{rgb}{0.5,0.0,0.5}
\definecolor{backcolour}{rgb}{0.97,0.97,0.97}

\lstdefinelanguage{json}{
    morestring=[b]",
    morecomment=[l]{//},
    morecomment=[s]{/*}{*/},
    stringstyle=\color{codegreen},
    commentstyle=\color{codegray},
    keywordstyle=\color{codeblue}\bfseries,
    morekeywords={true,false,null},
    literate={\ }{{\ }}1 
}

\usepackage{threeparttable}
\usepackage{xcolor,pifont}
\usepackage{url}
\usepackage{hyperref, epsfig, endnotes}
\usepackage{makecell}
\usepackage[normalem]{ulem}
\usepackage[misc]{ifsym}
\usepackage{subcaption}
\usepackage{cleveref}
\usepackage{tcolorbox}
\usepackage{python}
\usepackage{enumitem}

\usepackage{adjustbox}
\definecolor{codegray}{rgb}{0.5,0.5,0.5}
\definecolor{codeblue}{rgb}{0.0,0.0,0.6}
\definecolor{codegreen}{rgb}{0.1,0.5,0.1}
\definecolor{codepurple}{rgb}{0.58,0,0.82}
\definecolor{backcolour}{rgb}{0.97,0.97,0.97}

\AtBeginDocument{%
  }

\definecolor{customblue}{HTML}{006ca6}
\definecolor{customgreen}{HTML}{009264}
\definecolor{custombrown}{HTML}{ff3d00}
\AtEndPreamble{

}

\newcommand{\ea}{\textit{et~al.}}

\begin{document}

\title{A Dual-Loop Agent Framework for Automated Vulnerability Reproduction}

\author{Bin Liu}
\email{liubin1999@stu.hit.edu.cn}
\affiliation{%
  \institution{Harbin Institute of Technology, Shenzhen}
  \country{China}
}

\author{Yanjie Zhao}
\authornotemark[1]
\email{yanjie\_zhao@hust.edu.cn}
\affiliation{%
  \institution{Huazhong University of Science and Technology}
  \country{China}
  }

\author{Zhenpeng Chen}
\email{zpchen@tsinghua.edu.cn}
\affiliation{%
  \institution{Tsinghua University}
  \country{China}
  }

\author{Guoai Xu}
\authornote{Yanjie Zhao (yanjie\_zhao@hust.edu.cn) and Guoai Xu (xga@hit.edu.cn) are the corresponding authors.}
\email{xga@hit.edu.cn}
\affiliation{%
  \institution{Harbin Institute of Technology, Shenzhen}
  \country{China}
}

\author{Haoyu Wang}
\email{haoyuwang@hust.edu.cn}
\affiliation{%
  \institution{Huazhong University of Science and Technology}
  \country{China}
}

\begin{abstract}
Automated vulnerability reproduction from CVE descriptions requires generating executable Proof-of-Concept (PoC) exploits and validating them in target environments. This process is critical in software security research and practice, yet remains time-consuming and demands specialized expertise when performed manually. While LLM agents show promise for automating this task, existing approaches often conflate exploring attack directions with fixing implementation details, which leads to unproductive debugging loops when reproduction fails.
To address this, we propose \tool, an LLM-based dual-loop agent framework following a plan-execute-evaluate paradigm. The Strategic Planner analyzes vulnerability semantics and target code to produce structured attack plans. The Tactical Executor generates PoC code and validates it through progressive verification. The Adaptive Refiner evaluates execution results and routes failures to different loops: the \textit{Tactical Loop} for code-level refinement, while the \textit{Strategic Loop} for attack strategy replanning. This dual-loop design enables the framework to escape ineffective debugging by matching remediation to failure type.
Evaluation on two benchmarks covering 617 real-world vulnerabilities demonstrates that \tool achieves 82.9\% and 54.3\% reproduction success rates on SecBench.js and PatchEval, respectively, outperforming the best baseline by 11.3\% and 20.4\%. Human evaluation confirms that generated PoCs achieve comparable code quality to human-written exploits in readability and reusability.
\end{abstract}



\maketitle

\section{Introduction}\label{sec:intro}

Software vulnerabilities pose an escalating threat to digital security. The National Vulnerability Database has documented explosive growth in vulnerability disclosures in recent years~\cite{nvd_statistics_2025}. However, available CVE reports mainly provide natural language descriptions or structured metadata, without the executable and reproducible Proof-of-Concept (PoC) exploits that are essential for verifying authenticity and assessing impact. Empirical studies show that successfully reproducing disclosed vulnerabilities remains challenging even for professional security teams due to inadequate information in reports~\cite{mu2018understanding}, creating substantial uncertainty in vulnerability verification, patch validation, and defense deployment.

\textbf{The scarcity of executable PoCs stems from the complexity of manual reproduction.} Transforming vulnerability descriptions into working PoCs requires understanding exploitation mechanisms from text information, configuring appropriate runtime environments for specific software versions, and implementing attack code that successfully triggers the vulnerability. These steps demand specialized security expertise and substantial time investment, making manual reproduction inherently difficult to scale as vulnerability disclosures continue to grow. Automated approaches are therefore essential to bridge this reproduction gap.

Traditional automated techniques, including static analysis~\cite{codeql,semgrep}, dynamic fuzzing~\cite{afl,aflplusplus}, and symbolic execution~\cite{klee,kleef2024}, cannot address this challenge, as they lack mechanisms to interpret natural language vulnerability specifications and generate corresponding exploitation code. Large Language Models (LLMs) present a qualitatively different approach. Their ability to understand and synthesize code enables direct transformation from textual vulnerability descriptions to executable programs~\cite{llmvulnsurvey2024}. This potential has motivated recent investigations into LLM-based agents augmented with execution environments and tool interfaces for autonomous PoC generation. Representative systems include OpenHands~\cite{wang2024openhands}, a general software engineering agent that can interact with code repositories and be adapted for vulnerability reproduction tasks; CAI~\cite{mayoralvilches2025cai}, an open-source framework enabling security professionals to construct AI-driven offensive automation; and PoCGen~\cite{simsek2025pocgen}, a specialized system targeting npm package vulnerabilities through multi-stage pipelines that combine static analysis with reasoning.

Although these systems represent important progress, \textbf{existing LLM-based agents face a fundamental architectural limitation}: their single-cycle workflows repeatedly revise generated code without distinguishing the underlying causes of failure. When a PoC fails to execute properly, the agent cannot identify whether the problem lies in choosing the wrong attack method or simply making mistakes while implementing a correct method. This inability to distinguish leads to either wasting resources on perfecting exploitation ideas that cannot succeed or discarding promising approaches due to easily fixable errors. These issues are amplified by insufficient examination of vulnerability semantics before generating code and unreliable validation techniques that cannot confidently assess whether the generated PoC actually triggers the target vulnerability.

To fill this gap, we introduce \tool, an agent framework that employs two distinct feedback mechanisms to separate strategic exploitation planning from tactical code implementation for automated vulnerability reproduction. The framework operates through dual loops. The \textit{Tactical Loop} iterates between Tactical Executor and Adaptive Refiner, correcting code-level issues while preserving the attack strategy. When tactical fixes prove insufficient, the \textit{Strategic Loop} returns control to Strategic Planner, enabling recovery from incorrect vulnerability interpretations. By decoupling strategy selection from code debugging, this design avoids both the problem of endlessly refining unworkable attack plans and the problem of abandoning valid plans due to fixable implementation mistakes.

Specifically, \tool includes three core modules. First, 
\textit{Strategic Planner} processes CVE descriptions to identify exploitation-critical details and inspects vulnerable codebases to construct structured attack plans, establishing a thorough understanding of the vulnerability before any code is written. Second, \textit{Tactical Executor} generates PoC code from attack plans and validates it through progressively stringent multi-layer checks, from syntactic validity to semantic correctness to execution behavior to differential testing, pinpointing exactly where exploitation attempts fail. Third, \textit{Adaptive Refiner} analyzes verification results to determine failure causes, either triggering replanning when the attack methodology is flawed or performing targeted refinement for implementation errors, while maintaining a sparse experience index for efficient knowledge reuse.

In summary, our contributions are as follows:
\begin{itemize}[leftmargin=*]
    \item We propose \tool, the first dual-loop agent framework separating strategic planning from tactical execution for automated vulnerability reproduction from CVE descriptions.
    \item We design three core modules with novel components: \textit{Strategic Planner} with structured attack planning, \textit{Tactical Executor} with progressive multi-layer verification, and \textit{Adaptive Refiner} with dual-dimension feedback and sparse experience retrieval.
    \item We evaluate \tool on two benchmarks: 387 npm vulnerabilities from SecBench.js~\cite{secbench} and 230 vulnerabilities across Go, JavaScript, and Python from PatchEval~\cite{patcheval}. \tool achieves 82.9\% and 54.3\% reproduction success rates, outperforming state-of-the-art baselines by 11.3\% and 20.4\% respectively. Human evaluation confirms that generated PoCs achieve comparable quality to human-written exploits, scoring 4.15 versus 3.62 on a 5-point scale with 14.6\% improvement in readability.
\end{itemize}

\section{Background}

\subsection{Problem Formalization}

We formalize the vulnerability reproduction problem as follows. Given a vulnerability report $V$ containing a natural language description $D$, an affected software package $P$, a vulnerable version $v$, and optional metadata $M$ (e.g., CVE identifiers, CVSS scores), the goal is to automatically generate an executable PoC exploit $E$ that satisfies three requirements:

\textbf{Triggering Requirement.} The PoC must successfully trigger the vulnerability when executed against the vulnerable version $P_{v}$, demonstrating the security impact described in $D$. This requires the exploit to satisfy all environmental preconditions and execute the precise attack sequence that activates the vulnerable code path.

\textbf{Validation Requirement.} The exploitation must be objectively verifiable through observable evidence, such as unauthorized access, data leakage, privilege escalation, denial of service, or code execution. Since many vulnerabilities do not produce crashes, verification cannot rely solely on abnormal program termination.

\textbf{Specificity Requirement.} The PoC should demonstrate vulnerability-specific behavior rather than generic system failures. It must fail when executed against a patched version $P_{v+1}$ or alternative implementations lacking the vulnerability, confirming that observed effects stem from the specific weakness rather than unrelated issues.

This problem is particularly challenging because it requires bridging multiple semantic gaps: from unstructured natural language to structured exploitation logic, from abstract vulnerability descriptions to concrete attack implementations, and from static code analysis to dynamic execution validation. Unlike vulnerability detection, which identifies potential weaknesses, or patch generation, which repairs code, reproduction demands demonstrating exploitability through working attacks.

\subsection{Traditional Vulnerability Analysis Approaches}

Tools like CodeQL~\cite{codeql_github}, Semgrep~\cite{semgrep}, and Infer~\cite{infer} systematically analyze program structure without execution, identifying potential vulnerabilities through pattern matching and dataflow analysis. While highly effective for detecting known vulnerability patterns, they cannot execute attacks or verify exploitability, thus limiting their applicability to vulnerability reproduction.

Fuzzing tools such as AFL~\cite{afl}, AFL++~\cite{aflplusplus}, and LibFuzzer~\cite{libfuzzer} generate test inputs to trigger crashes and errors. Symbolic execution tools like KLEE~\cite{klee} and KLEEF~\cite{kleef2024} systematically explore program paths. However, these techniques require executable binaries or source code with known vulnerable states and cannot interpret natural language vulnerability descriptions.

Systems like AEG~\cite{avgerinos2011aeg} and Mayhem~\cite{cha2012mayhem} generate exploits from program crashes or vulnerabilities identified through other means. While pioneering in automation, they assume pre-identified vulnerable locations and focus on memory corruption vulnerabilities, struggling with logic flaws common in web applications.

All traditional approaches share a critical constraint: they operate on program artifacts with known vulnerable states but cannot comprehend the natural language CVE descriptions that characterize real-world vulnerability disclosures. This semantic gap prevents direct application to vulnerability reproduction, where the only available information is often unstructured text.

\subsection{Security-Focused LLM-Based Agent Systems}

Recent research has adapted agent architectures to penetration testing and exploitation. PentestGPT~\cite{deng2023pentestgpt} provides interactive attack guidance by combining LLM reasoning with security tool invocation. AutoPentest~\cite{henke2025autopentest} automates reconnaissance workflows through structured agent collaboration. Fang ~\ea~\cite{fang2024llmagentsautonomouslyexploit} demonstrated that GPT-4 can exploit 87 of one-day vulnerabilities when provided with CVE descriptions. Zhu ~\ea~\cite{zhu2025teamsllmagentsexploit} showed that coordinated agent teams outperform individual agents on complex multi-step exploits.

\noindent \textbf{Automated PoC Generation Systems.} Several systems specifically target automated PoC generation from vulnerability descriptions. CVE-GENIE~\cite{ullah2025cvegenie} coordinates specialized agents through understanding, planning, and execution phases. PoCGen~\cite{simsek2025pocgen} combines LLMs with static and dynamic program analysis through a multi-phase pipeline, reaching 77\% success on npm packages. VulnBot~\cite{wang2024vulnbot} employs multi-agent collaboration for comprehensive vulnerability assessment. PoCo~\cite{poco2025} generates smart contract exploits compatible with testing frameworks.

\subsection{Limitations of Existing Approaches}\label{sec:limitations}

Despite progress in LLM-based security agents, existing systems exhibit three critical limitations that motivate our work:

\textbf{(C1) Lack of Structured Pre-Analysis.} Current PoC generation methods jump directly from vulnerability descriptions to code synthesis without building comprehensive models of exploitation requirements. For example, PoCGen~\cite{simsek2025pocgen} focuses its analysis phase on locating vulnerable code rather than systematically modeling how the vulnerability can be triggered. CVE-GENIE~\cite{ullah2025cvegenie} noted that this absence of structured pre-analysis causes agents to resort to blind trial-and-error when initial attempts fail, wasting computational resources on fundamentally flawed attack strategies. The problem intensifies for logic vulnerabilities requiring precise sequences of operations, where understanding triggering conditions before implementation is critical.

\textbf{(C2) Unreliable Verification Mechanisms.} Determining whether a PoC successfully triggers a vulnerability is inherently challenging. Many exploits produce no crashes: cross-site scripting may silently inject scripts, SQL injection may return normal responses while leaking data, and authentication bypasses may appear as ordinary logins. Current approaches employ one of three inadequate strategies: (1) predefined security checks that cover only limited vulnerability patterns, failing on novel or complex exploits; (2) generic execution feedback analyzing return codes or output text without vulnerability-specific oracles, leading to false positives from coincidental output matches; or (3) differential testing comparing behavior between vulnerable and patched versions, which requires access to both versions and struggles when patches are unavailable or introduce unrelated changes. Research on LLM-based code generation has demonstrated that superficial execution success often masks deeper functional failures~\cite{liu2023evalplus}, a problem exacerbated in security contexts where correct execution may still fail to exploit the vulnerability.

\textbf{(C3) Flat Decision-Making Without Failure Diagnosis.} Existing agent systems employ single-loop generate-execute-refine cycles that treat all failures uniformly~\cite{shinn2023reflexion}. When a PoC fails, they cannot distinguish whether the failure stems from a fundamentally incorrect attack strategy or merely implementation bugs in an otherwise sound approach. This conflation of strategic and tactical errors leads to two failure modes: (1) wasted effort on iterative refinements of flawed approaches that cannot possibly succeed, such as attempting SQL injection on endpoints lacking database queries; (2) premature abandonment of promising strategies due to minor implementation issues, such as syntax errors or incorrect parameter formats. Experience utilization further complicates matters, with systems either ignoring past attempts entirely, losing valuable insights, or injecting full execution histories into context, causing token inflation that degrades model performance~\cite{memgpt2024}.

\textbf{Our Approach.} \tool addresses these limitations through three architectural mechanisms that directly correspond to the identified challenges. To address \textbf{C1}, we introduce a dual-loop architecture that separates strategic planning from tactical execution, ensuring structured vulnerability analysis produces comprehensive attack models before code generation. To address \textbf{C2}, we design progressive multi-layer verification that combines symbolic analysis with execution feedback and differential testing, providing reliable assessment of exploitation success across diverse vulnerability types. To address \textbf{C3}, we implement adaptive failure diagnosis with sparse experience indexing, enabling intelligent routing between strategic replanning and tactical refinement while maintaining efficient context utilization. The following section details our framework design.
\section{Methodology}
\label{sec:method}

\subsection{Overview}
\label{sec:overview}
\begin{figure*}[htbp]
  \centering
  \includegraphics[width=\linewidth]{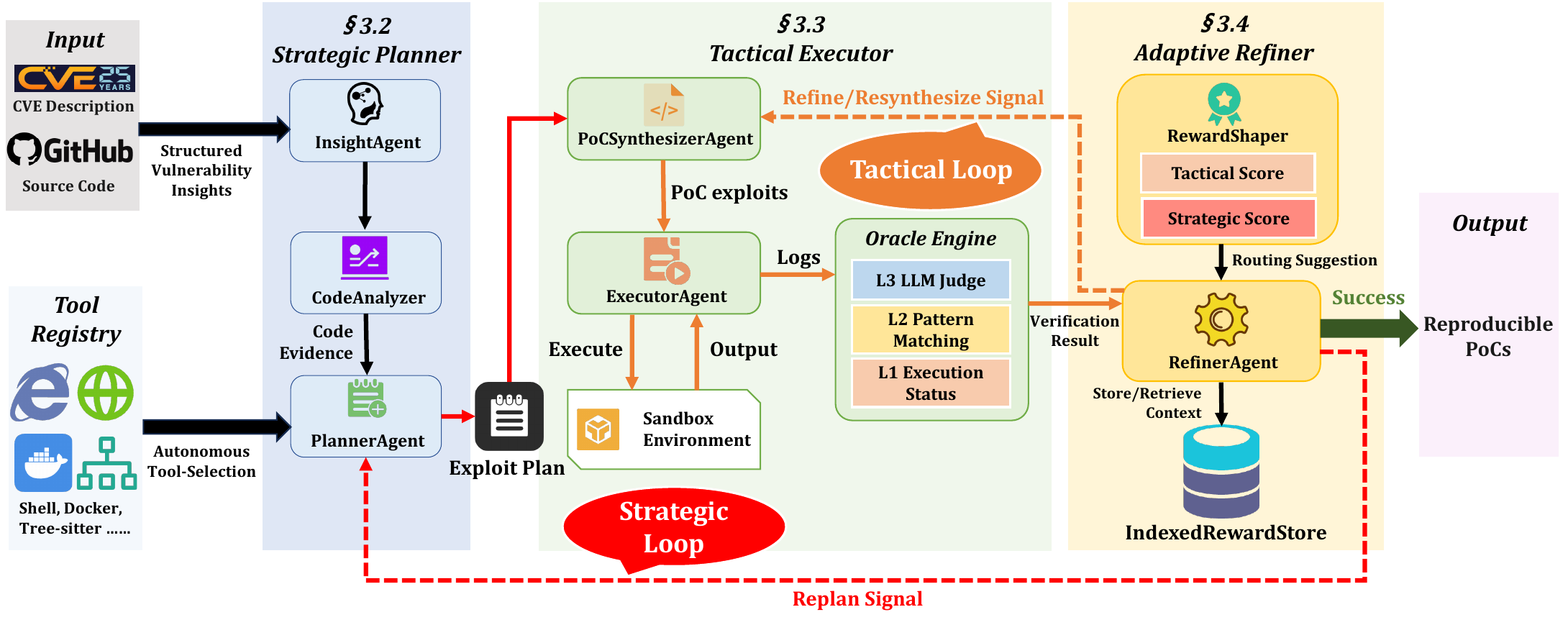}
  \caption{Overview of the \tool framework.}
  \label{fig:architecture}
\end{figure*}

As illustrated in \autoref{fig:architecture}, \tool is a dual-loop agent comprising three core modules: Strategic Planner, Tactical Executor, and Adaptive Refiner. \textbf{\textit{Strategic Planner}} tackles the lack of structured understanding by analyzing CVE descriptions and target codebases to synthesize attack plans before code generation. \textbf{\textit{Tactical Executor}} addresses unreliable verification through progressive multi-layer checks that pinpoint specific failure points rather than producing binary outcomes. \textbf{\textit{Adaptive Refiner}} resolves flat decision-making by diagnosing failure causes and routing to appropriate correction mechanisms: tactical refinement for implementation errors or strategic replanning for fundamental approach failures.

In particular, \tool operates through two feedback loops. The \textcolor{orange}{\textbf{Tactical Loop}} iterates between \textit{\textbf{Tactical Executor}} and \textit{\textbf{Adaptive Refiner}}, correcting code-level issues while preserving the attack strategy. When tactical fixes prove insufficient, the \textcolor{red}{\textbf{Strategic Loop}} returns control to \textbf{\textit{Strategic Planner}}, enabling recovery from incorrect vulnerability interpretations. This separation prevents wasted effort on refining fundamentally flawed approaches while avoiding premature abandonment of sound strategies marred by minor bugs.

\subsection{Strategic Planner}

Existing methods lack structured vulnerability understanding (\textbf{C1} in \autoref{sec:limitations}), causing blind trial-and-error when initial attempts fail. \textbf{The core insight is that effective exploitation requires bridging three semantic gaps before code generation:} translating high-level descriptions into precise attack vectors (semantic gap), locating specific vulnerable code paths amid large codebases (contextual gap), and selecting appropriate execution strategies for diverse targets (operational gap). We design three specialized agents that bridge these gaps through structured analysis.

\subsubsection{InsightAgent: Semantic Gap Bridging}

CVE descriptions contain rich exploitation information encoded in unstructured natural language. The \texttt{InsightAgent} employs schema-guided prompting to extract this information into a structured specification that constrains subsequent analysis. Given a CVE description, the agent identifies vulnerability type (e.g., code injection, path traversal), attack vector (HTTP, command-line interface), vulnerable parameters, and key APIs involved in exploitation. This structured insight serves two purposes: it provides a high-level constraint on what to look for in the codebase, and it generates targeted search queries combining vulnerability patterns with identified APIs.

\autoref{lst:insight} shows an example extraction where the agent identifies that user input flows to an \texttt{eval} call via the \texttt{template} parameter on the \texttt{/api/render} endpoint. This specification eliminates ambiguity and guides targeted code analysis.

\subsubsection{CodeAnalyzer: Contextual Gap Bridging}

Triggering vulnerabilities requires locating vulnerable code paths and understanding their invocation requirements. The \texttt{CodeAnalyzer} combines static analysis with heuristic matching to collect this evidence efficiently. It first runs Semgrep security rules to identify candidate locations with severity scores. It then filters findings by matching against vulnerability-specific keywords from the \texttt{InsightAgent} output, focusing on high-confidence matches. For matched locations, the analyzer applies pattern recognition to extract concrete exploitation parameters: endpoint URLs from route annotations, parameter names from input retrieval patterns, and payload hints from the vulnerable operations.

\autoref{lst:code_evidence} illustrates the extracted evidence, showing the exact file, line number, endpoint, parameter name, and a suggested payload structure. This provides the concrete code context needed to construct valid PoCs.

\begin{figure}[t]
\centering
\begin{lstlisting}[caption={InsightAgent Output},label={lst:insight},basicstyle=\ttfamily\footnotesize,xleftmargin=8pt,xrightmargin=8pt]
{
  "vulnerability_type": "Code Injection",
  "attack_vector": "HTTP",
  "vulnerable_params": ["template", "userInput"],
  "key_apis": ["eval", "Function"],
  "entry_points": ["/api/render", "/api/compile"],
  "summary": "User-controlled input flows into eval without sanitization"
}
\end{lstlisting}

\vspace{0.5em}

\begin{lstlisting}[caption={CodeAnalyzer Output},label={lst:code_evidence},basicstyle=\ttfamily\footnotesize,xleftmargin=8pt,xrightmargin=8pt]
{
  "file": "src/template.js",
  "line": 42,
  "endpoint": "/api/render",
  "parameter": "template",
  "payload_hint": "${require('child_process').exec('whoami')}",
  "confidence": 0.85
}
\end{lstlisting}
\caption{Structured outputs from Strategic Planner agents.}
\label{fig:agent_output}
\end{figure}

\subsubsection{PlannerAgent: Operational Gap Bridging}

Different targets require distinct execution strategies: npm vulnerabilities need direct function invocation within Node.js runtimes, while containerized applications require Docker-based execution. Hard-coded rules are brittle and fail to generalize across vulnerability types. \textbf{We employ LLM-based autonomous planning where the model selects appropriate tools based on target characteristics.}

The system maintains a \texttt{ToolRegistry} (\autoref{tab:tool_registry}) where each tool is registered with capability descriptions. Given vulnerability context (target type, vulnerability type, code characteristics) and available tool descriptions, the LLM generates a structured exploitation plan containing execution steps, malicious payloads derived from the vulnerability type, and oracle configuration specifying success criteria. This approach handles edge cases that predefined rules miss while ensuring principled tool selection through capability-aware reasoning.

\begin{table}[h]
\centering
\caption{Tool Registry with capability descriptions.}
\label{tab:tool_registry}
\resizebox{0.7\linewidth}{!}{%
\begin{tabular}{@{}lll@{}}
\toprule
\textbf{Tool} & \textbf{Capability} & \textbf{Primary Use Cases} \\ \midrule
\texttt{node\_executor} & Execute JavaScript locally & npm vulnerabilities \\
\texttt{docker\_executor} & Isolated container execution & Multi-language targets \\
\texttt{semgrep} & Security-focused static analysis & Vulnerability pattern detection \\
\texttt{http} & Network request generation & API endpoint testing \\
\texttt{tree-sitter} & Syntax tree analysis & Function signature extraction \\
\bottomrule
\end{tabular}%
}
\end{table}

\subsection{Tactical Executor}

Determining whether a PoC successfully exploits a vulnerability remains challenging, as many exploits produce no crashes or obvious anomalies (\textbf{C2} in \autoref{sec:limitations}). \textbf{Our key insight is that verification confidence can be progressively built through multiple complementary evidence layers.} Rather than relying on single-point checks that conflate execution success with exploitation success, we design three components that transform plans into executable PoCs and validate them through increasingly stringent verification: from syntactic correctness to semantic alignment to behavioral evidence to differential confirmation.

\subsubsection{PoCSynthesizerAgent: Plan-Guided Code Generation}

The \texttt{PoCSynthesizerAgent} generates exploit code from attack plans and vulnerable source code analysis. Unlike template-based approaches that match predefined patterns, LLM-based synthesis adapts to arbitrary function signatures, module structures, and vulnerability trigger conditions by understanding code semantics.

The synthesis process begins with automatic language detection based on syntactic features (e.g., \texttt{require()} for JavaScript, \texttt{def} for Python). The agent then analyzes vulnerable function behavior to understand whether it executes commands, validates strings, or writes files. \textbf{This analysis prevents common mistakes like attempting command injection against functions performing only string validation.} The agent constructs payloads tailored to vulnerability types: shell metacharacters with observable side effects for command injection, \texttt{\_\_proto\_\_} manipulation for prototype pollution, \texttt{../} sequences for path traversal. Generated PoCs embed predefined success markers (e.g., \texttt{VULNERABILITY\_TRIGGERED}) that enable reliable verification.

\subsubsection{ExecutorAgent: Unified Multi-Environment Execution}

The \texttt{ExecutorAgent} provides environment-agnostic execution through automatic backend selection. For npm vulnerabilities, it uses \texttt{node\_executor} with proper dependency resolution. For PatchEval targets, it uses \texttt{docker\_executor} with automatic interpreter detection and compilation support for statically-typed languages. The executor captures artifacts including stdout, stderr, exit codes, and execution duration. These structured results enable both verification and detailed failure diagnosis.

\subsubsection{OracleEngine: Progressive Multi-Layer Verification}

\autoref{sec:limitations} noted that simple pattern matching produces false positives, while LLM-based judgment is prohibitively expensive. \textbf{Our key insight is that most failures can be quickly rejected through fast checks, reserving expensive semantic reasoning for ambiguous cases.} The \texttt{OracleEngine} implements three progressively stringent verification layers with confidence-driven escalation.

\textbf{L1 Execution Status} performs fast rule-based verification examining exit codes, timeout conditions, and fatal error patterns (e.g., \texttt{SyntaxError}, \texttt{ModuleNotFoundError}). Scripts exhibiting these failures are immediately rejected. Scripts completing with exit code 0 proceed to pattern matching.

\textbf{L2 Pattern Matching} detects predefined success markers (e.g., \texttt{POC\_SUCCESS}, \texttt{EXPLOIT\_SUCCESS}) and vulnerability-specific evidence patterns (e.g., file creation confirmations, unauthorized data access). This layer provides high-confidence positive signals without LLM invocation.

\textbf{L3 LLM Judge} handles ambiguous cases where pattern matching is inconclusive. The LLM receives execution output with vulnerability context and determines whether evidence convincingly demonstrates successful reproduction. This layered design achieves both accuracy and efficiency: most executions are resolved by fast checks, while expensive reasoning applies only when necessary.

\subsection{Adaptive Refiner}
\label{sec:ad}

When a PoC fails, existing systems cannot distinguish whether the failure stems from fundamentally incorrect attack strategies or merely implementation bugs (\textbf{C3} in \autoref{sec:limitations}). \textbf{We observe that execution failures decompose into orthogonal dimensions requiring different remediation strategies.} The Adaptive Refiner addresses this through three mechanisms: structured feedback interpretation diagnoses failure root causes, dual-dimension routing directs failures to appropriate correction loops, and sparse experience indexing enables knowledge reuse without context inflation.

\subsubsection{RewardShaper: Dual-Dimension Failure Diagnosis}

The \texttt{RewardShaper} transforms execution results into structured feedback signals that enable principled routing decisions. Rather than treating all failures uniformly, the system decomposes execution quality along two orthogonal axes.

The \textcolor{orange}{\textit{Tactical Score}} $s_{tactical} \in [-1, 1]$ quantifies code-level correctness: syntax validity, runtime stability, and parameter accuracy. High tactical scores indicate the code executes correctly but may target the wrong attack vector; low scores indicate implementation bugs requiring code fixes.

The \textcolor{red}{\textit{Strategic Score}} $s_{strategic} \in [0, 10]$ measures overall attack strategy soundness: progress toward vulnerability reproduction and alignment with target characteristics. High strategic scores indicate the approach is fundamentally correct but needs execution refinement. Low strategic scores indicate the attack vector is fundamentally wrong and requires replanning.

\textbf{This dual-dimension decomposition enables targeted correction:} tactical issues trigger code refinement within the current strategy, while strategic issues trigger replanning. The scoring mechanism employs LLM-as-Judge with explicit evaluation rubrics (\autoref{tab:reward_criteria}). This approach captures nuanced failure modes that deterministic heuristics miss, enabling more precise diagnostic feedback.

\begin{table}[htbp]
\centering
\caption{Dual-dimension reward criteria and routing logic.}
\label{tab:reward_criteria}
\resizebox{\columnwidth}{!}{%
\begin{tabular}{@{}clp{0.7\linewidth}l@{}}
\toprule
\textbf{Dimension} & \textbf{Score} & \textbf{Failure or Reward Mode} & \textbf{Routing} \\ 
\midrule
\multirow{3}{*}{Tactical} 
 & $-1.0$ & Fatal errors: SyntaxError, ImportError, ModuleNotFoundError & Resynthesize \\
 & $-0.5$ & Runtime crashes: TypeError, ReferenceError, unhandled exceptions & Refine \\
 & $1.0$ & Clean execution: exit code 0, no stderr output & (Strategic) \\ 
\midrule
\multirow{4}{*}{Strategic} 
 & $0$--$2$ & No meaningful interaction: immediate crash, wrong target & Replan \\
 & $3$--$5$ & Weak exploitation signals: target reached, wrong parameters & Refine \\
 & $6$--$8$ & Strong exploitation signals: partial success, stack trace exposed & Refine \\
 & $10$ & Successful reproduction: success marker detected, vulnerability confirmed & Success \\ 
\bottomrule
\end{tabular}%
}
\end{table}

Based on dual-dimension scores, the \texttt{RewardShaper} produces preliminary routing suggestions with four distinct outcomes: \textit{success} when both dimensions indicate successful exploitation, \textit{resynthesize} when critical code errors require complete regeneration, \textit{refine} when localized modifications suffice, and \textit{replan} when the fundamental strategy requires revision. These correspond to the dual-loop architecture where \textit{refine} and \textit{resynthesize} trigger the \textcolor{orange}{\textbf{Tactical Loop}}, while \textit{replan} triggers the \textcolor{red}{\textbf{Strategic Loop}}.

\subsubsection{RefinerAgent: Adaptive Routing and Targeted Modification}

The \texttt{RefinerAgent} serves as the decision hub, validating and potentially overriding routing suggestions based on deeper failure context analysis. \textbf{For instance, repeated \textit{refine} attempts without progress trigger escalation to \textit{replan} even when scores suggest otherwise, preventing infinite tactical loops.}

When the routing decision is \textit{refine}, the agent performs targeted tactical modifications while preserving the attack strategy. It classifies failure root causes into diagnostic categories (\texttt{wrong\_tool}, \texttt{syntax\_error}, \texttt{env\_missing}) and modifies payloads, parameters, or configurations accordingly. For \textit{resynthesize} decisions, it routes to the \texttt{PoCSynthesizerAgent} for complete code regeneration. For \textit{replan} decisions, it triggers the \textcolor{red}{\textbf{Strategic Loop}} by returning control to the \texttt{PlannerAgent}.

\subsubsection{IndexedRewardStore: Sparse Experience Retrieval}

\autoref{sec:limitations} noted that naive experience injection causes context inflation while stateless approaches repeat mistakes. \textbf{Our key insight is that most historical attempts are irrelevant to current failures, and efficient retrieval requires indexing by failure characteristics.}

The \texttt{IndexedRewardStore} maintains compact reward records containing iteration numbers, dual-dimension scores, state tags, improvement hints, and metadata (error types, tools used). Each record receives a unique identifier enabling prompts to reference past attempts through IDs rather than embedding full content. During context construction, a two-phase retrieval strategy operates: retrieving the $K$ most recent rewards for temporal continuity, and querying rewards filtered by error type ranked by strategic score for relevant patterns. Actual content loads on demand with in-memory caching for frequently accessed records.

This design achieves efficient experience utilization: relevant past attempts inform current decisions without overwhelming context, enabling the system to learn from failures while maintaining bounded prompt sizes. The sparse indexing mechanism proves critical for handling complex vulnerabilities requiring numerous iterative refinements.
\section{Experimental Setup}

\subsection{Research Questions}\label{sec:setup_rq}

We evaluate the effectiveness of \tool by comparing it with state-of-the-art approaches for automated vulnerability reproduction and PoC generation. We focus on the following research questions (RQs):

\textbf{RQ1:} How effective is \tool compared with existing approaches in generating executable and verifiable vulnerability PoCs?

\textbf{RQ2:} What are the contributions of the three core modules, \textbf{\textit{Strategic Planner}}, \textbf{\textit{Tactical Executor}}, and \textbf{\textit{Adaptive Refiner}}, to the reproduction success rate?

\textbf{RQ3:} What factors affect \tool's reproduction performance across different conditions?

\textbf{RQ4:} How does the code quality of \tool-generated PoCs compare to human-written ground truth PoCs?

To answer RQ1, we compare three representative baselines, including PoCGen~\cite{simsek2025pocgen}, OpenHands~\cite{wang2024openhands}, and CAI~\cite{mayoralvilches2025cai}, all using the same backbone LLM to ensure fair comparison of system designs.
To answer RQ2, we conduct ablation studies aligned with our three-module architecture, performing module-level ablations for \textbf{\textit{Strategic Planner}} and \textbf{\textit{Adaptive Refiner}}, and mechanism-level ablation for the Progressive Multi-Layer Oracle within \textbf{\textit{Tactical Executor}}.
To answer RQ3, we analyze how various factors affect reproduction performance, including backbone LLM choice, vulnerability type, CVE publication year, description length, and code complexity.
To answer RQ4, we compare the code quality of \tool-generated PoCs against human-written ground truth PoCs from both datasets through human expert evaluation, assessing three dimensions: Readability, Reusability, and Professionalism.

\subsection{Datasets}\label{sec:setup_dataset}

We evaluate \tool on two real-world vulnerability reproduction benchmarks that collectively cover a wide range of vulnerability types and programming ecosystems.

\textbf{SecBench.js.} 
SecBench.js~\cite{secbench} is an executable security benchmark suite for server-side JavaScript vulnerabilities. The original dataset contains 600 vulnerabilities covering five common vulnerability classes in the npm ecosystem. From this dataset, we select 387 vulnerabilities that have official CVE identifiers, ensuring well-documented vulnerability descriptions and expected exploitation behaviors. Each entry contains a CVE identifier and detailed vulnerability description, vulnerable npm package name and version, source code repository with commit information, executable exploit payload, and oracle for validating successful exploitation. The selected samples span diverse vulnerability types including Remote Code Execution, Prototype Pollution, Command Injection, Path Traversal, and Regular Expression Denial of Service.

\textbf{PatchEval.}
PatchEval~\cite{patcheval} is a multi-language vulnerability benchmark, originally designed for patch evaluation but equally suitable for reproduction due to its Dockerized sandbox environments with PoC tests. The benchmark comprises 1,000 vulnerabilities sourced from CVEs reported between 2015 and 2025, covering 65 distinct CWE categories across Go, JavaScript, and Python. We use the reproducible subset of 230 CVEs that are equipped with Dockerized sandbox environments enabling runtime validation through both PoC tests and unit tests. Each entry contains the CVE identifier and description, CWE information, vulnerable and fixed code snippets, patch diffs, and test commands for reproduction.

\subsection{Baselines}\label{sec:setup_baseline}

We compare \tool against three representative approaches for automated vulnerability reproduction and PoC generation.

\textbf{PoCGen}~\cite{simsek2025pocgen} is a state-of-the-art approach for generating PoC exploits for npm package vulnerabilities. It combines large language models with static and dynamic analyses to extract code evidence, generate candidate exploits, and validate whether the exploit reproduces the vulnerability.

\textbf{OpenHands}~\cite{wang2024openhands} is an open platform for building AI software developers as generalist agents. We adapt OpenHands to our vulnerability reproduction task by providing CVE descriptions and source code context, configuring it to iteratively generate and execute candidate PoCs.

\textbf{CAI}~\cite{mayoralvilches2025cai} is a cybersecurity-oriented agent framework designed for offensive security workflows including reconnaissance, exploitation, and PoC generation. We configure CAI with the same execution environment and verification interface as \tool to ensure fair comparison.

\subsection{Evaluation Metrics}\label{sec:setup_metric}

Following previous work on automated software testing~\cite{liu2023evalplus,chen2021humaneval}, we employ multiple metrics to evaluate effectiveness and efficiency.

\textbf{Effectiveness Metrics.}
Reproduction Success Rate measures the percentage of vulnerabilities for which the system successfully generates an executable PoC that triggers the expected vulnerability behavior. A PoC is considered successful if it executes without errors, triggers the vulnerability verified by our oracle, and completes within the time budget.

\textbf{Efficiency Metrics.}
Average Execution Time measures the end-to-end time from CVE input to final PoC generation.
Token Consumption measures the total tokens processed by the backbone LLM during the reproduction process.

\subsection{Implementation Details}\label{sec:setup_impl}

We implement \tool in Python 3.11 and manage dependencies using Conda. All experiments are conducted on a workstation equipped with an Intel Core i9-12900K CPU, an NVIDIA RTX 3090 GPU with 24GB memory, 32GB DDR4 RAM, and 2TB NVMe storage. To ensure safety and reproducibility, all PoC executions are sandboxed in lightweight Docker containers with strict resource limits and network isolation. We set a hard timeout of 600 seconds per CVE and limit the refinement iterations to 3 strategic and 5 tactical cycles.

For RQ1 and RQ2, we use DeepSeek-V3.2~\cite{deepseek2025v3} as the backbone LLM for both \tool and all baselines, ensuring fair comparison of system designs. All models are configured with temperature 0.1 and maximum token limit of 128K.

For RQ3, we conduct impact analysis using a stratified sample of 100 CVEs. We evaluate three LLM backbones including GPT-5.2~\cite{openai2025gpt5} and two open-source models (Qwen3-Coder-30B~\cite{qwen2025coder} and GPT-OSS-20B~\cite{openai2025gptoss}) deployed locally using Ollama. We also analyze the impact of vulnerability type, CVE publication year, description length, and code complexity on reproduction performance.
\section{Experimental Results}\label{sec:exp_result}

This section presents our experimental results and answers the four RQs outlined in \autoref{sec:setup_rq}.

\subsection{RQ1: Comparison with Baselines}

To answer RQ1, we compare the performance of \tool with three state-of-the-art baselines: PoCGen~\cite{simsek2025pocgen}, OpenHands~\cite{wang2024openhands}, and CAI~\cite{mayoralvilches2025cai}. 
\autoref{tab:rq1_main} presents the comprehensive comparison including overall Reproduction Success Rate, Token Consumption, and detailed breakdown by vulnerability types and programming languages. We make four key observations from these results:

(1) \textbf{\tool achieves the highest reproduction success rate across both datasets.} 
As shown in \autoref{tab:rq1_main}, \tool achieves an RSR of \textbf{82.9\%} on SecBench.js, outperforming the best baseline CAI by \textbf{11.3\%} absolute improvement.
On PatchEval, \tool demonstrates strong multi-language generalization with an RSR of \textbf{54.3\%}, surpassing CAI by 20.4\% and OpenHands by 32.1\%.
PoCGen, which is specialized for npm vulnerabilities, achieves only 7.8\% on SecBench.js and cannot generalize to PatchEval at all.

(2) \textbf{\tool demonstrates superior performance across diverse vulnerability types.}
\tool achieves the highest success rate in 4 out of 5 vulnerability categories on SecBench.js.
For Prototype Pollution, \tool achieves 96.0\% compared to 74.8\% of CAI and 72.8\% of OpenHands.
For Code Injection, \tool achieves 90.0\% compared to 55.0\% of CAI and 45.0\% of OpenHands.
For ReDoS vulnerabilities, \tool achieves 89.5\% while OpenHands drops to only 14.0\%.
The only category where CAI outperforms \tool is Path Traversal (71.2\% vs 46.2\%), likely due to its specialized security tooling.

(3) \textbf{\tool exhibits robust cross-language generalization capability.}
On PatchEval, \tool maintains consistently high performance across Python, Go, and JavaScript ecosystems.
\tool achieves 52.9\% on Python vulnerabilities, doubling the 25.7\% of CAI and nearly quadrupling the 14.3\% of OpenHands.
For Go vulnerabilities, \tool achieves 47.0\% compared to 24.1\% of OpenHands and 16.9\% of CAI.
This demonstrates the effectiveness of our autonomous planning and multi-environment execution in adapting to different language ecosystems.

(4) \textbf{\tool achieves higher efficiency with lower token consumption.}
\tool consumes only \textbf{6.1M tokens} on SecBench.js, which is \textbf{16$\times$ fewer} than the 96.0M tokens of CAI while achieving higher success rate.
On PatchEval, \tool uses only \textbf{10.0M tokens} compared to 98.6M of CAI and 101.3M of OpenHands, representing a \textbf{10$\times$ reduction}.
This efficiency stems from our progressive multi-layer oracle that avoids expensive LLM verification when simpler checks suffice, and our sparse experience indexing that prevents redundant context injection.

\begin{table*}[t]
\centering
\caption{Comprehensive comparison of \tool with baselines on SecBench.js and PatchEval benchmarks. RSR denotes Reproduction Success Rate. The breakdown shows performance across vulnerability types and programming languages.}
\label{tab:rq1_main}
\resizebox{\textwidth}{!}{%
\begin{tabular}{l cc ccccc cc ccc}
\toprule
\multirow{3}{*}{\textbf{Method}} & \multicolumn{7}{c}{\textbf{SecBench.js (387 CVEs)}} & \multicolumn{5}{c}{\textbf{PatchEval (230 CVEs)}} \\
\cmidrule(lr){2-8} \cmidrule(lr){9-13}
 & \multicolumn{2}{c}{\textbf{Overall}} & \multicolumn{5}{c}{\textbf{By Vulnerability Type (\%)}} & \multicolumn{2}{c}{\textbf{Overall}} & \multicolumn{3}{c}{\textbf{By Language (\%)}} \\
\cmidrule(lr){2-3} \cmidrule(lr){4-8} \cmidrule(lr){9-10} \cmidrule(lr){11-13}
 & \textbf{RSR (\%)} & \textbf{Tokens (M)} & \textbf{PP} & \textbf{CI} & \textbf{CodeI} & \textbf{ReDoS} & \textbf{PT} & \textbf{RSR (\%)} & \textbf{Tokens (M)} & \textbf{Python} & \textbf{Go} & \textbf{JS} \\ 
\midrule
PoCGen~\cite{simsek2025pocgen} & 7.8 & 13.2 & 0.0 & 0.0 & 0.0 & 0.0 & 37.5 & - & - & - & - & - \\
OpenHands~\cite{wang2024openhands} & 55.8 & 17.3 & 72.8 & 70.9 & 45.0 & 14.0 & 41.2 & 22.2 & 101.3 & 14.3 & 24.1 & 27.3 \\
CAI~\cite{mayoralvilches2025cai} & 71.6 & 96.0 & 74.8 & 84.8 & 55.0 & 50.9 & \textbf{71.2} & 33.9 & 98.6 & 25.7 & 16.9 & 59.7 \\
\midrule
\textbf{\tool} & \textbf{82.9} & \textbf{6.1} & \textbf{96.0} & \textbf{88.6} & \textbf{90.0} & \textbf{89.5} & 46.2 & \textbf{54.3} & \textbf{10.0} & \textbf{52.9} & \textbf{47.0} & \textbf{63.6} \\ 
\bottomrule
\end{tabular}%
}

\begin{minipage}{\textwidth}
\footnotesize
\textit{Note: PP = Prototype Pollution (151), CI = Command Injection (79), CodeI = Code Injection (20), ReDoS (57), PT = Path Traversal (80), JS = JavaScript. Best results are in \textbf{bold}.}
\end{minipage}
\end{table*}

\finding{1}{\tool achieves 82.9\% RSR on SecBench.js and 54.3\% on PatchEval, outperforming the best baseline CAI by 11.3\% and 20.4\% respectively.}

\subsection{RQ2: Module Contribution Analysis}

To answer RQ2, we conduct ablation studies aligned with our three-module architecture to investigate the contribution of each module to the reproduction success rate.
Following our methodology design, we perform module-level ablations for \textbf{\textit{Strategic Planner}} and \textbf{\textit{Adaptive Refiner}}, and mechanism-level ablation for the LLM Oracle within \textbf{\textit{Tactical Executor}}.
\autoref{tab:rq2_ablation} presents the ablation study results on SecBench.js.

\subsubsection{\textbf{\textit{Strategic Planner}}}
To investigate the impact of strategic planning, we deploy a variant that removes the \textbf{\textit{Strategic Planner}} module entirely (w/o \textbf{\textit{Strategic Planner}}).
In this variant, the system directly generates PoC code from CVE descriptions without producing structured vulnerability insights, code evidence, or exploitation plans.
As shown in \autoref{tab:rq2_ablation}, removing the \textbf{\textit{Strategic Planner}} results in a substantial drop in RSR from 82.9\% to 53.7\%, representing a 29.2\% absolute decrease.
Analysis of failure cases reveals a fundamental shift in failure patterns: while the full system primarily fails due to code-level bugs that can be iteratively fixed, the ablated variant predominantly fails due to strategy-level errors such as targeting wrong attack vectors, misunderstanding vulnerability mechanisms, or selecting inappropriate exploitation techniques.
This demonstrates that \textbf{\textit{Strategic Planner}} is the most critical module because it provides high-level vulnerability semantics understanding, structured attack strategies, and concrete exploitation blueprints that guide downstream code generation.
Without these strategic foundations, the system cannot effectively navigate the complex search space of vulnerability reproduction.

\subsubsection{\textbf{\textit{Adaptive Refiner}}}
To explore the impact of adaptive refinement and memory mechanisms, we deploy a variant that removes the \textbf{\textit{Adaptive Refiner}} module (w/o \textbf{\textit{Adaptive Refiner}}).
In this variant, the system uses fixed-iteration naive retry without reward-driven routing, dual-dimension scoring, or sparse experience indexing.
As shown in \autoref{tab:rq2_ablation}, the variant without \textbf{\textit{Adaptive Refiner}} experiences a 23.2\% decrease in RSR, dropping from 82.9\% to 59.7\%.
Failure analysis reveals that this variant frequently becomes trapped in ineffective debugging loops, repeatedly attempting similar fixes for fundamentally flawed strategies.
The \textbf{\textit{Adaptive Refiner}} is essential because it maintains historical execution trajectories with both \textcolor{orange}{\textit{Tactical Score}} (code-level feedback) and \textcolor{red}{\textit{Strategic Score}} (exploitation-level feedback), enabling the system to distinguish between fixable code errors and fundamental strategy failures.
This dual-dimension feedback mechanism guides intelligent routing decisions between local refinement and global replanning.

\subsubsection{LLM Oracle}
To evaluate the contribution of LLM-based verification within the \textbf{\textit{Tactical Executor}}, we deploy a variant that uses only static keyword matching for success verification (w/o LLM Oracle).
As shown in \autoref{tab:rq2_ablation}, removing the LLM Oracle results in a 7.7\% decrease in RSR, from 82.9\% to 75.2\%, and notably consumes 1.7$\times$ more tokens.
This demonstrates that static verification performs worse while costing more.
The underlying reason is that static keyword matching cannot accurately interpret diverse execution outputs across different vulnerability types.
Without the LLM Oracle's contextual understanding, the system frequently misclassifies successful reproductions as failures or vice versa, leading to unnecessary retry iterations and wasted computational resources.
In contrast, the LLM Oracle dynamically adapts its judgment criteria based on the specific vulnerability context, providing accurate feedback signals that enable efficient convergence.

\begin{table}[t]
\centering
\caption{Ablation study results on SecBench.js (387 CVEs).}
\label{tab:rq2_ablation}
\small
\begin{tabular}{l c c c c}
\toprule
\textbf{Variant} & \textbf{RSR (\%)} & \textbf{$\Delta$RSR} & \textbf{Time (s)} & \textbf{Tokens (M)} \\ \midrule
\tool (Full) & \textbf{82.9} & - & 132.7 & 6.1 \\
w/o \textbf{\textit{Strategic Planner}} & 53.7 & -29.2\% & \textbf{46.2} & \textbf{2.9} \\
w/o \textbf{\textit{Adaptive Refiner}} & 59.7 & -23.2\% & 79.2 & 6.1 \\
w/o LLM Oracle & 75.2 & -7.7\% & 123.9 & 10.5 \\ \bottomrule
\end{tabular}
\end{table}

\finding{2}{All three modules are essential, with \textbf{\textit{Strategic Planner}} contributing most (29.2\%) and \textbf{\textit{Adaptive Refiner}} enabling feedback routing (23.2\%).}

\subsection{RQ3: Impact Analysis}

To answer RQ3, we conduct an impact analysis examining how various factors affect \tool's reproduction performance.
We investigate five key dimensions: backbone LLM choice, vulnerability type, CVE publication year, description length, and code complexity.
We compare three backbone configurations: GPT-5.2~\cite{openai2025gpt5} as a cloud-based commercial model, Qwen3-Coder-30B~\cite{qwen2025coder} (code-specialized), and GPT-OSS-20B~\cite{openai2025gptoss} (general-purpose) deployed locally.
All experiments are conducted on a stratified sample of 100 CVEs from SecBench.js (20 per vulnerability type) with identical system configurations.

\subsubsection{Overall Performance}
GPT-5.2 achieves the highest RSR of 77.0\%, followed by Qwen3-Coder-30B at 68.0\% and GPT-OSS-20B at 66.0\%.
This demonstrates that \tool's dual-loop architecture exhibits robust cross-backbone generalization, with open-source models reaching 88.3\% of the cloud model's performance, enabling practical local deployment for security-sensitive scenarios.
Notably, Qwen3-Coder-30B outperforms GPT-OSS-20B by 2\%, confirming that code-specialized training benefits PoC generation tasks.
However, efficiency differs substantially: GPT-5.2 completes in 104.5s on average, while Qwen3-Coder-30B requires 587.6s (5.6$\times$ slower) and GPT-OSS-20B requires 1038.5s (10$\times$ slower).

\subsubsection{Performance by Vulnerability Type}
As shown in \autoref{fig:rq3_radar}, distinct performance patterns emerge across vulnerability types.
For Prototype Pollution and Command Injection, all models achieve high success rates (80-95\%), with GPT-5.2 leading.
Interestingly, for Code Injection, both open-source models (90\%) outperform GPT-5.2 (85\%), suggesting that code-specialized training provides advantages for certain vulnerability patterns.
ReDoS exhibits the largest performance gap: GPT-5.2 achieves 75\% while GPT-OSS-20B drops to only 35\%, indicating that complex regex analysis requires stronger reasoning capabilities.
Path Traversal remains challenging for all models (35-40\%), representing a common limitation that requires future improvement.

\subsubsection{Impact of Task Complexity}
\begin{figure*}[t]
\centering
\scalebox{0.9}{%
\begin{minipage}{\textwidth}
\centering
\subfigure[By Vulnerability Type]{
\includegraphics[width=0.45\textwidth]{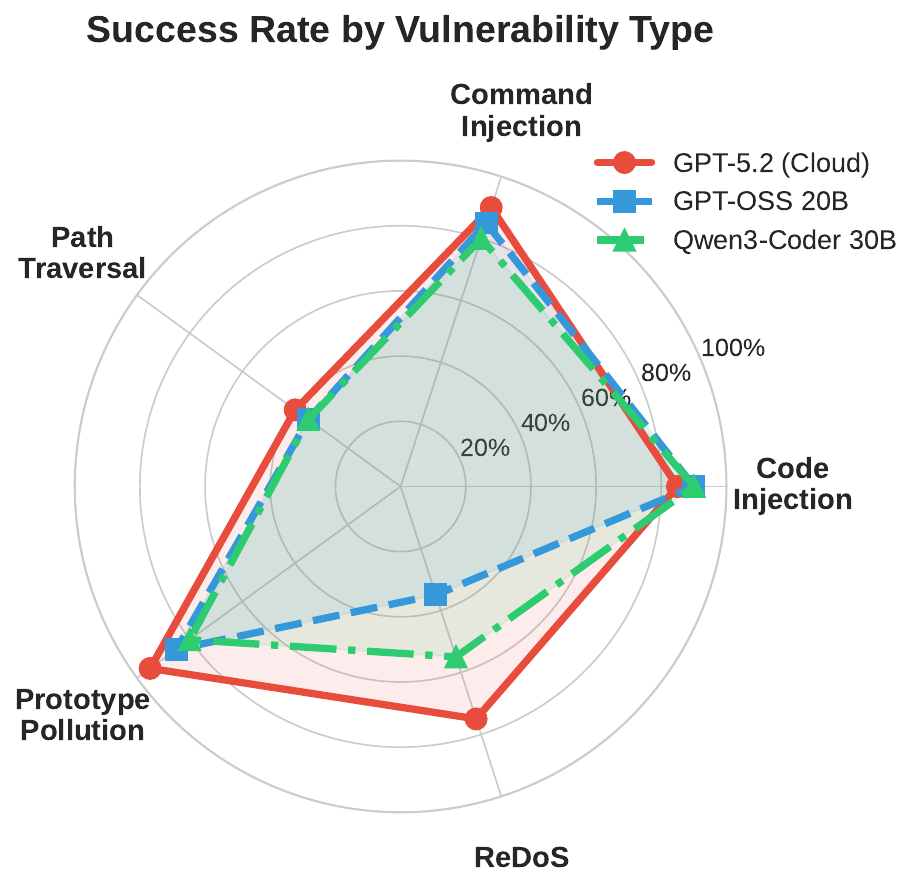}
\label{fig:rq3_radar}
}%
\hfill
\subfigure[By CVE Year]{
\includegraphics[width=0.45\textwidth]{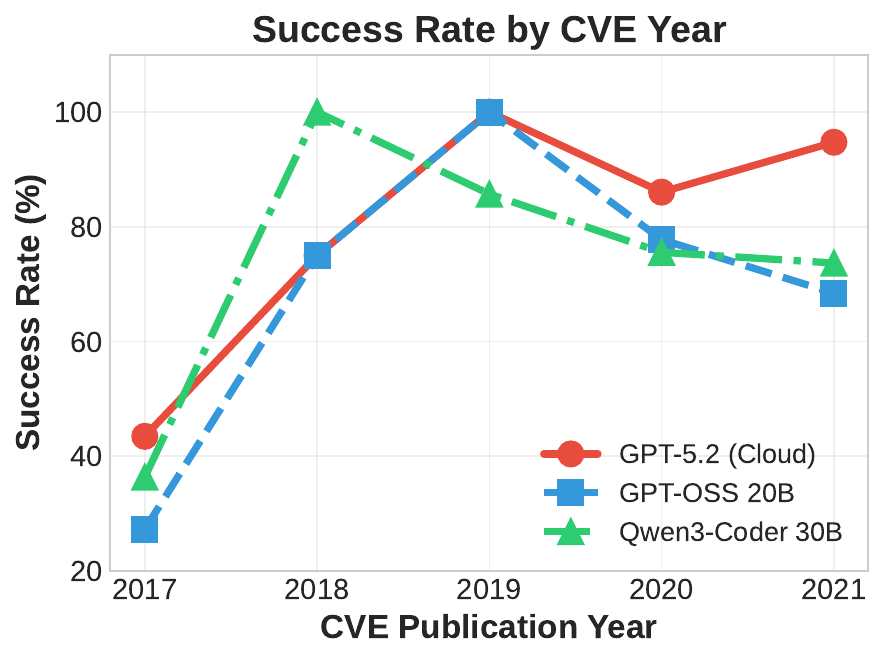}
\label{fig:rq3_year}
}
\\
\subfigure[By Description Length]{
\includegraphics[width=0.45\textwidth]{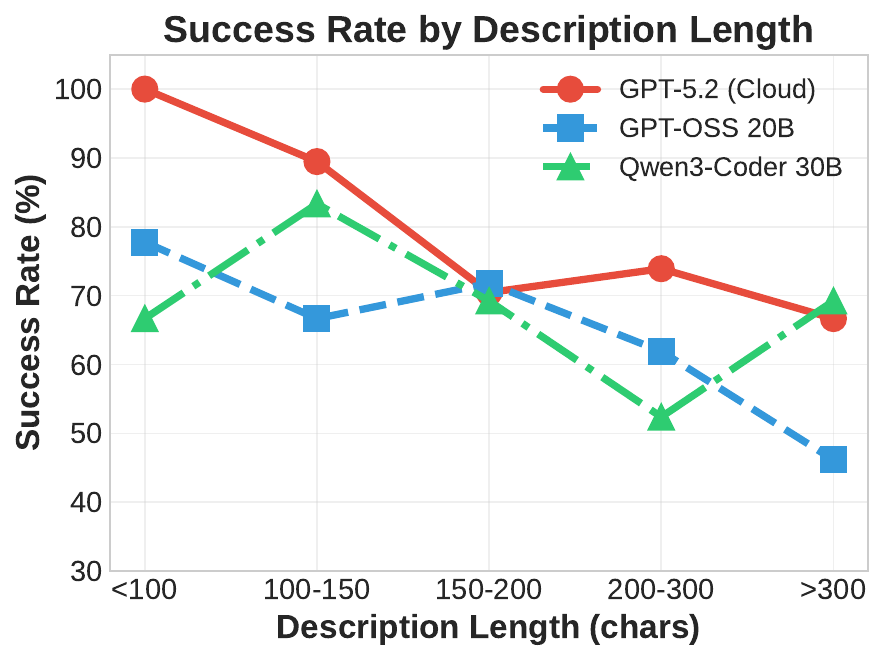}
\label{fig:rq3_desc}
}%
\hfill
\subfigure[By Code Complexity]{
\includegraphics[width=0.45\textwidth]{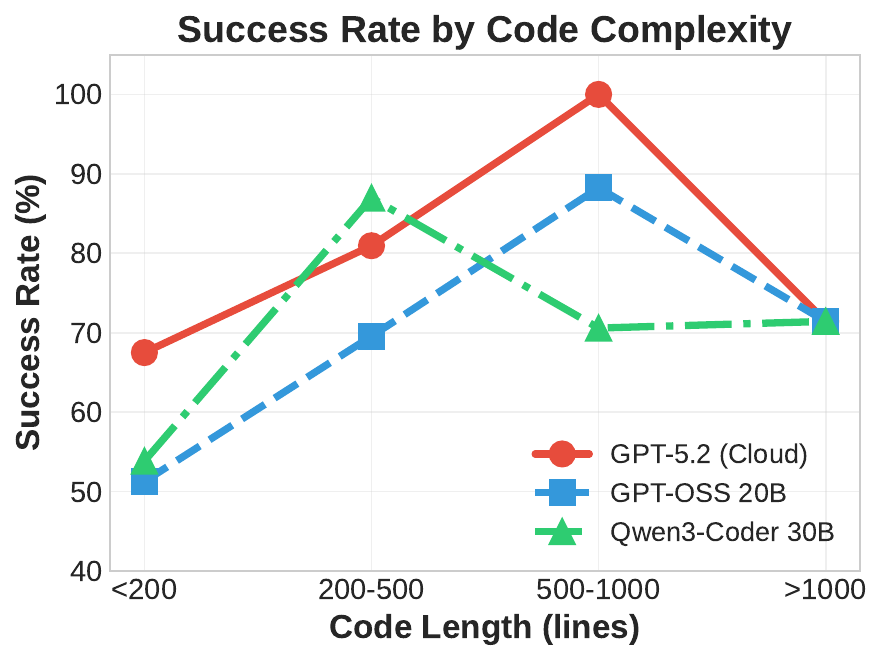}
\label{fig:rq3_code}
}
\end{minipage}%
}
\caption{Impact analysis across different dimensions. (a) Performance varies by vulnerability type, with open-source models surpassing cloud on Code Injection. (b) Older CVEs are harder to reproduce. (c) Shorter descriptions favor cloud models. (d) Medium-sized codebases yield optimal results.}
\label{fig:rq3_all}
\end{figure*}

We further analyze how CVE characteristics affect model performance across three dimensions.
As shown in \autoref{fig:rq3_year}, older vulnerabilities (2017) prove more challenging with success rates of 27-44\%, while recent CVEs (2019-2021) achieve 75-100\%.
This temporal effect likely reflects both evolving vulnerability patterns and training data coverage.
As shown in \autoref{fig:rq3_desc}, GPT-5.2 excels on short descriptions (<100 chars) with 100\% success, while open-source models show greater degradation as description length increases.
As shown in \autoref{fig:rq3_code}, medium-sized codebases (500-1000 lines) yield the highest success rates, while both very small (<200 lines) and very large (>1000 lines) codebases present challenges.

\finding{3}{Reproduction performance varies by LLM backbone, vulnerability type, CVE year, and task complexity. Notably, open-source models achieve 88\% of cloud performance, demonstrating architectural robustness and enabling practical local deployment.}

\subsection{RQ4: Code Quality Evaluation}

To answer RQ4, we compare the code quality of \tool-generated PoCs against human-written ground truth PoCs through human expert evaluation.

\noindent\textbf{Evaluation Setup.} From the successful reproductions in RQ1, we randomly select 100 vulnerabilities (50 from SecBench.js and 50 from PatchEval). For each vulnerability, we collect both the PoC generated by \tool and the ground truth PoC provided in the dataset, yielding 200 PoCs in total. To eliminate potential bias from recognizing generation patterns, we mix all 200 PoCs and randomize their presentation order.

We evaluate three dimensions of code quality using a 1-to-5 Likert scale (5 for excellent, 4 for good, 3 for acceptable, 2 for marginal, 1 for poor): (1) \textbf{Readability}, whether the code is well-structured with clear variable naming and logical flow; (2) \textbf{Reusability}, whether the code is parameterized, configurable, and easily adaptable for similar vulnerabilities; (3) \textbf{Professionalism}, whether the code follows security research best practices with appropriate comments, error handling, and clear exploitation logic.

We recruit three cybersecurity professionals for human evaluation: two penetration testing engineers with five years of industry experience and one PhD student in cybersecurity with four years of research experience. First, the three evaluators complete a pilot training session in which they jointly discuss 10 example PoCs (which are not included in the 100 vulnerabilities used for evaluation) to establish a shared understanding of the scoring criteria. Then, each evaluator independently scores individual PoCs without knowing their source, ensuring absolute rather than comparative ratings. For each sample, evaluators receive: (1) the CVE description, (2) the relevant source code of the vulnerable software, and (3) the PoC code to be evaluated.

\noindent\textbf{Results.} Each evaluator completes the evaluation in around 4 hours, resulting in a total of 1,800 individual ratings (3 evaluators $\times$ 200 PoCs $\times$ 3 dimensions). For each PoC, we compute the score of each dimension by averaging across all three evaluators. Since each dataset contains 50 CVEs with two PoCs per CVE (one ground truth and one generated), the final score for each dataset-source combination is computed by averaging across the corresponding 50 PoCs. \autoref{tab:rq4_quality} presents the human evaluation results.

On SecBench.js, \tool-generated PoCs achieve higher average scores than ground truth (3.91 vs 3.60). The improvement is most pronounced in Reusability, where \tool scores 3.87 compared to 3.21 for ground truth, representing a 20.6\% improvement. This is because generated PoCs tend to be more parameterized and configurable. On PatchEval, \tool also achieves higher average scores than ground truth (3.86 vs 3.75). The advantage is primarily in Readability (3.96 vs 3.74) and Reusability (3.76 vs 3.38), while ground truth maintains a slight edge in Professionalism (4.12 vs 3.85) due to more sophisticated error handling and domain-specific edge case coverage.

Notably, \tool outperforms ground truth in Readability across both datasets, with 14.6\% improvement on SecBench.js (4.15 vs 3.62) and 5.9\% on PatchEval (3.96 vs 3.74), indicating that LLM-generated code tends to have cleaner structure and more consistent naming conventions.

\begin{table}[htbp]
\centering
\caption{Code quality comparison between \tool-generated and ground truth PoCs. All scores are on a 1-5 Likert scale based on human expert evaluation.}
\label{tab:rq4_quality}
\resizebox{0.6\linewidth}{!}{%
\begin{tabular}{ll cccc}
\toprule
\textbf{Dataset} & \textbf{Source} & \textbf{Read.} & \textbf{Reus.} & \textbf{Prof.} & \textbf{Avg.} \\ 
\midrule
\multirow{2}{*}{SecBench.js} & Ground Truth & 3.62 & 3.21 & \textbf{3.98} & 3.60 \\
 & \tool & \textbf{4.15} & \textbf{3.87} & 3.72 & \textbf{3.91} \\
\midrule
\multirow{2}{*}{PatchEval} & Ground Truth & 3.74 & 3.38 & \textbf{4.12} & 3.75 \\
 & \tool & \textbf{3.96} & \textbf{3.76} & 3.85 & \textbf{3.86} \\
\bottomrule
\end{tabular}%
}
\end{table}

\finding{4}{\tool-generated PoCs achieve comparable code quality to human-written ground truth, with consistent advantages in Readability and Reusability.}

\section{Discussion}
\label{sec:discussion}

\subsection{Why Does \tool Work?}

We attribute the effectiveness of \tool to two key mechanisms, illustrated through a representative case study.

\textbf{Observation 1: Reward-driven routing enables adaptive recovery through three distinct pathways.}
The \textbf{\textit{Adaptive Refiner}} uses dual-dimension rewards (\textcolor{orange}{\textit{Tactical Score}} and \textcolor{red}{\textit{Strategic Score}}) to determine the recovery route: \textit{refine} for near-success cases, \textit{resynthesize} for code-level issues, and \textit{replan} for strategy-level failures.

We take CVE-2020-28273 as an example to illustrate this mechanism. This vulnerability is a prototype pollution in the \texttt{set-in} npm package, where user-controlled paths can pollute \texttt{Object.prototype}. As shown in \autoref{tab:trajectory_case}, the system went through 5 iterations with all three routing decisions before successfully reproducing the vulnerability.

\begin{table*}[htbp]
\centering
\caption{Execution trajectory for CVE-2020-28273 (Prototype Pollution in \texttt{set-in} package).}
\label{tab:trajectory_case}
\resizebox{\textwidth}{!}{%
\begin{tabular}{c l c c l l l}
\toprule
\textbf{Iter} & \textbf{State} & \textbf{Tac.} & \textbf{Str.} & \textbf{Feedback} & \textbf{Improvement Hints} & \textbf{Route} \\
\midrule
0 & partial\_exec & 0.0 & 8.0 & 
\begin{tabular}[t]{@{}l@{}}Payload failed but reached vuln point.\\ Output: \texttt{"pollution not achieved"}\end{tabular} & 
\begin{tabular}[t]{@{}l@{}}Adjust payload format.\\ Try \texttt{constructor.prototype} path.\end{tabular} & 
refine \\
\midrule
1 & module\_error & $-1.0$ & 0.0 & 
\begin{tabular}[t]{@{}l@{}}\texttt{Cannot find module 'set-in'}.\\ PoC cannot start.\end{tabular} & 
\begin{tabular}[t]{@{}l@{}}Ensure module installed.\\ Check script paths.\end{tabular} & 
resynth \\
\midrule
2 & no\_progress & 0.0 & 0.0 & 
\begin{tabular}[t]{@{}l@{}}Steps skipped (KnownFailure).\\ Exit code: None.\end{tabular} & 
\begin{tabular}[t]{@{}l@{}}Resolve env config.\\ Verify package path.\end{tabular} & 
replan \\
\midrule
3 & no\_progress & 0.0 & 0.0 & 
\begin{tabular}[t]{@{}l@{}}Consecutive failures detected.\\ System triggers fallback.\end{tabular} & 
\begin{tabular}[t]{@{}l@{}}Switch to Node executor.\\ Use pre-configured env.\end{tabular} & 
replan \\
\midrule
4 & exploited & 0.8 & 8.0 & 
\begin{tabular}[t]{@{}l@{}}Node execution succeeded!\\ \texttt{Object.prototype.polluted="yes"}\end{tabular} & 
-- & 
success \\
\bottomrule
\end{tabular}%
}
\end{table*}

In Iteration 0, the code reached the vulnerability point but the payload failed to pollute the prototype. The high \textcolor{red}{\textit{Strategic Score}} (8.0) indicated near-success, so the system chose \textit{refine} to adjust the payload format. Iteration 1 encountered a module loading error (\textcolor{orange}{\textit{Tactical Score}}=$-1.0$), triggering \textit{resynthesize} to regenerate the PoC. Iterations 2-3 showed no progress due to environment issues, leading to consecutive \textit{replan} decisions that ultimately triggered the fallback mechanism. In Iteration 4, the system switched to a pre-configured Node environment and successfully confirmed the prototype pollution.

\textbf{Observation 2: Sparse experience indexing reduces token consumption while maintaining effectiveness.}
The \texttt{IndexedRewardStore} in \autoref{sec:ad} retrieves relevant past experiences based on vulnerability type and error patterns, injecting historical insights into the planning process to avoid redundant exploration. \autoref{tab:memory_effect} evaluates the impact of experience indexing.

\begin{table}[h]
\centering
\footnotesize
\caption{Effect of sparse experience indexing.}
\label{tab:memory_effect}
\resizebox{0.8\linewidth}{!}{%
\begin{tabular}{lcccc}
\toprule
Configuration & RSR (\%) & Tokens (M) & Avg Iters & Avg Time (s) \\
\midrule
\tool (full) & 82.9 & 6.1 & 1.8 & 132 \\
w/o Experience Indexing & 72.1 ($-$10.8) & 9.9 ($+$62\%) & 2.9 ($+$61\%) & 198 ($+$50\%) \\
\bottomrule
\end{tabular}
}%
\end{table}

When experience indexing is disabled, the reproduction success rate drops by 10.8\% points, while token consumption increases by 62\% and average execution time increases from 132 to 198 seconds. The average number of iterations also increases from 1.8 to 2.9, indicating that the system requires more attempts to converge. These results demonstrate that sparse experience indexing significantly accelerates convergence by providing relevant historical context.

\subsection{Key Differences from Existing Methods}

\autoref{tab:method_diff} compares \tool with the baseline methods introduced in \autoref{sec:setup_baseline}. \tool is the only method that provides strategic planning through dual-loop architecture, enabling replanning when tactical approaches fail. It is also unique in offering dual-dimension reward signals for fine-grained routing decisions (refine/resynthesize/replan) and sparse experience indexing that reduces token consumption by 62\% while avoiding context inflation. Additionally, \tool supports dynamic executor selection to handle environment-specific issues across multiple languages.

\begin{table}[h]
\centering
\caption{Comparison of \tool with existing methods.}
\label{tab:method_diff}
\resizebox{0.7\linewidth}{!}{
\begin{tabular}{lcccc}
\toprule
Capability & PoCGen & CAI & OpenHands & \tool \\
\midrule
Strategic Planning & \texttimes & \texttimes & \texttimes & \checkmark \\
Code Analysis & Static & Limited & LLM & LLM \\
Sandbox Execution & Limited & Docker & Docker & Dynamic \\
Feedback Signal & None & Binary & Binary & Dual-dim \\
Route Decision & None & None & None & 3 routes \\
Experience Reuse & None & Context & Context & Sparse Index \\
Language Support & npm & Multi & Multi & Multi \\
\bottomrule
\end{tabular}
}
\end{table}

\section{Threats to Validity}
\label{sec:threats}
\noindent \textbf{Baseline Selection.}
We compare \tool against three baseline methods: PoCGen, CAI, and OpenHands. PoCGen is specifically designed for npm package vulnerabilities, CAI is a security-focused LLM agent, and OpenHands is a general-purpose software engineering agent. These baselines may not fully represent all existing vulnerability reproduction approaches, and future work will extend comparisons to additional tools.

\noindent \textbf{Dataset Selection.}
Our evaluation uses SecBench.js and PatchEval, which cover JavaScript, Python, and Go ecosystems. The results may not fully generalize to other languages like C/C++ or Java, which have different build systems and memory models. The reliance on existing benchmarks also limits coverage to documented CVEs with available source code.

\noindent \textbf{Metric Selection.}
Reproduction Success Rate measures functional correctness but does not capture PoC quality or minimality. A generated PoC may successfully trigger the vulnerability but contain redundant code or lack readability. While RQ4 provides additional quality evaluation, our metrics may still not fully reflect practical usefulness in real-world security assessments.

\section{Related Work}
\label{sec:related}

\indent\textbf{LLMs for Security.} LLMs have transformed security research by enabling comprehension of natural language vulnerability descriptions. Foundation models~\cite{feng2020codebert,guo2021graphcodebert,codellama} and security-specific variants~\cite{hanif2022vulberta,fu2022linevul} support vulnerability classification and localization. LLM-based detection~\cite{sun2024gptscan,sun2024llm4vuln} and repair systems~\cite{fu2022vulrepair,zhang2024autocoderover} demonstrate semantic understanding but focus on static analysis without execution feedback. Agent architectures~\cite{yao2023react,shinn2023reflexion,wei2022cot,wang2023planandsolve} enable iterative reasoning, while multi-agent systems~\cite{hong2023metagpt,sweagent2024,wang2024openhands} and memory mechanisms~\cite{memgpt2024,zhao2024expel} extend these ideas to complex software tasks.

\noindent\textbf{Automated PoC Generation.} Several systems target PoC generation from vulnerability descriptions. PoCGen~\cite{simsek2025pocgen} combines LLMs with program analysis in npm ecosystem, while CVE-GENIE~\cite{ullah2025cvegenie} coordinates multi-agent collaboration through resource gathering, environment reconstruction, and exploit generation phases. Recent work explores multi-agent teams for complex exploits~\cite{zhu2025teamsllmagentsexploit} and modular LLM frameworks with analysis-generation-verification pipelines~\cite{peng2025pwngpt}. Process-based feedback~\cite{zhao2024repair,arxiv2025aprsurvey} enables iterative refinement. OpenHands~\cite{wang2024openhands} provides an agent adaptable to software engineering tasks, while CAI~\cite{mayoralvilches2025cai} offers security-focused tooling. However, these approaches lack structured pre-analysis, employ unreliable verification, and use flat decision-making that conflates strategic and tactical errors~\cite{liu2023evalplus,shinn2023reflexion}. \tool addresses these through dual-loop architecture with progressive verification and sparse experience indexing.

\section{Conclusion}
\label{sec:conclusion}

In this paper, we introduced \tool, a dual-loop agent framework for automated PoC generation from CVE descriptions. \tool consists of three modules: \textbf{\textit{Strategic Planner}} for vulnerability analysis and attack planning, \textbf{\textit{Tactical Executor}} for PoC synthesis and progressive verification, and \textbf{\textit{Adaptive Refiner}} for reward-driven feedback routing and sparse experience indexing. Experimental results on 617 real-world vulnerabilities demonstrate that \tool achieves 82.9\% reproduction success rate on SecBench.js and 54.3\% on PatchEval, outperforming the best baseline by 11.3\% and 20.4\% respectively while consuming significantly fewer tokens. Impact analysis reveals robust performance across different LLM backbones, vulnerability types, and task complexities. Human evaluation confirms that generated PoCs achieve comparable code quality to human-written exploits in readability and reusability.

\bibliographystyle{ACM-Reference-Format}
\bibliography{main}

\end{document}